\documentclass{aa}
\usepackage{graphicx}
\usepackage{txfonts}
\begin{document}
\title{Broad iron emission lines in Seyfert Galaxies - re-condensation
  of gas onto an inner disk below the ADAF?}


\author{E. Meyer-Hofmeister \inst{1} and F. Meyer\inst{1}}
\offprints{Emmi Meyer-Hofmeister; emm@mpa-garching.mpg.de}
\institute
    {Max-Planck-Institut f\"ur Astrophysik, Karl-
     Schwarzschild-Str.~1, D-85740 Garching, Germany\\
     \mail{emm@mpa-garching.mpg.de}
}
\date{Received: / Accepted:}
\abstract
   {The number of strong iron $K_\alpha$ line detections in Seyfert AGN is
   clearly growing in the {\it{Chandra, XMM-Newton}} 
    \rm{and} {\it{Suzaku}} \rm{era}. 
   The iron emission lines are broad, some are
   relativistically blurred. These relativistic disk lines have also been   
   observed for galactic black hole X-ray binaries. Thermal
   components found in hard spectra were interpreted as an indication for
   a weak inner cool accretion disk underneath a hot corona.}
    {Accretion in low-mass X-ray binaries (LMXB) occurs during
      phases of high and low mass accretion rate, outburst and quiescence, 
      soft and hard spectral state, respectively. After the soft/hard 
      transition for some sources 
      a thermal component is found, which can be interpreted
      as sustained by re-condensation of gas from an 
      advection-dominated flow (ADAF) onto the disk. In view of the
      similarity of accretion flows around stellar mass and
      supermassive black holes we discuss whether the
      broad iron emission lines in Seyfert 1 AGN (Active Galactic Nuclei) can be
      understood as arising from a similar accretion flow geometry as
      in X-ray binaries.}
     {We derive accretion rates for those Seyfert galaxies for which broad iron 
      emission lines were observed, the ``best candidates'' in the 
      investigations of Miller (2007) and Nandra et al. (2007). For
      the evaluation of the Eddington-scaled rates we use the observed
      X-ray luminosity, bolometric corrections and black hole masses
      from the literature most values taken from the investigation of
      Fabian \& Vasudevan (2009).}
     {The accretion rates derived for the Seyfert galaxies in our
     sample are less than 0.1 of the Eddington rate for more than
     half of the sources. For $10^7$ to $10^8 M_\odot$ black holes in 
     Seyfert 1 AGN this limit corresponds to 0.01 to 0.2
     $M_\odot/{\rm{yr}}$. This documents that the sources probably
     are in a hard spectral state and iron emission lines can arise 
     from an inner weak accretion disk surrounded by an ADAF as
     predicted by the re-condensation model.
    Some of the remaining sources with higher accretion rates may be in a
    spectral state that is comparable to the ``very high'' state of LMXBs.}
    {Our investigation shows that in quite a number of Seyfert AGN
    the broad iron emission lines may indeed originate in a weak inner
    disk below the ADAF, close to the black hole,
    indicating the same accretion flow geometry as recently found
    for LMXBs. For the accretion history one then concludes that the
    accretion rates were higher in the outer radii at some earlier time
    .}


\titlerunning {Inner disks in Seyfert Galaxies}

\maketitle
%

\section{Introduction}
The physical processes of accretion onto galactic and supermassive
black holes and the accretion flow geometry in the innermost regions are 
key features for modeling relativistic spectral lines and for estimates of
the black hole spin. Observations in recent years with the {\it{Chandra
  X-ray Observatory}}, the {\it{X-ray Multi-Mirror Mission-Newton}} and
{\it{Suzuka}} reveal for some AGN broad, often relativistic iron
emission lines from the innermost regions close to the black hole. 
Iron emission lines are the most obvious response of an accretion disk 
to irradiation by an external source of hard X-rays. When X-rays
from a hot coronal flow fall on optically thick cool material,
they induce fluorescence and are backscattered, resulting in a Compton 
reflection spectrum with a prominent emission line of iron
K$\alpha$. The accretion geometry strongly depends on the mass accretion
rate. For LMXBs it is well known that on the one hand
for high rates an optically thick, geometrically thin,
radiatively efficient Shakura-Sunyaev accretion disk reaches inward to
the innermost stable circular orbit (ISCO), or on the other hand for low
accretion rates (lower than the critical rate of spectral state
transition $\dot
m_{\rm{crit}}$), a more spherical optically thin, hot advection-dominated 
flow fills the inner region (ADAF; Narayan \& Yi 1994, 1995,
Abramowicz et al. 1995; see Narayan 2005, Yuan 2007, Narayan \& McClintock 2008
for reviews). At larger distance an outer Shakura-Sunyaev disk exists,
whose truncation radius recedes as  the mass accretion rate
decreases. These ``high'' and ``low'' states are
documented by a soft and hard spectrum. The critical accretion rate $\dot
m_{\rm{crit}}$ lies around a few to 10 percent of the Eddington 
accretion rate (Esin et al. 1997). 

For supermassive black holes, especially low-luminosity AGN, truncated thin
disks were found, along with a hard spectrum (Narayan et
al. 1998). The power-law spectrum arising from
the coronal emission is scale-invariant (accretion rate in units of 
Eddington accretion rate), while the disk emission of AGN peaks in
the UV range, in contrast to the emission in soft X-rays from disks
around stellar mass black holes. Vasudevan \& Fabian (2007)
discuss the similarity of accretion states in AGN and galactic black 
hole sources and point out that the radiation in the
UV is an important contribution to the bolometric luminosity. They
derived bolometric corrections for AGN using recent results from
the {\it{ Far Ultraviolet Spectroscopic  Explorer (FUSE)}}, together
with data from the Optical Monitor (OM) archives and X-ray data from the 
{\it{XMM-Newton}} archive. They expanded the investigation and for the
first time took into account simultaneous observations (Vasudevan \&
Fabian 2009).

The accretion geometry in the two spectral states, either an ADAF (or
one of its variants) in the inner regions or a disk
reaching inward to the ISCO, seemed to 
exclude each other for a
long time. But the recent observations of LMXBs seem to indicate 
that during in intermediate state, the brightest hard state (after 
soft/hard transition in 
outburst decay), an ADAF in the inner region and a weak innermost
disk  can both be present at the same time (see Figure 1).
These co-existing hot and cool gas flows inward toward the black hole 
 clearly interact. The interaction causes mass
and angular momentum exchange between disk and ADAF, either a mass
flow from the disk into the ADAF, evaporation of the disk, or a reverse
mass flow, re-condensation of gas from the ADAF onto the disk (Liu et
al. 2006, Meyer et al. 2007).

Indications for a disk near the black hole during a canonical low-hard
state were found for
several X-ray binaries, with clearest hints in GX 339-4, SWIFT J1753.5-0127
and XTE J1817-330. The re-condensation model allows us to understand the
observed mass flow rates in the inner disk (Liu et al. 2007, Taam et
al. 2008). Observational evidence for the possible presence
of a thermal component during the hard state in eight sources, broad
skewed Fe K$\alpha$ lines in half of the sample, were shown in the recent work
by Reis et al. (2010). The authors interprete the reflection features 
as caused by illumination of a more-or-less permanent disk by the hard,
power-law component of a jet.

We note that an extremely skewed relativistic Fe
K$\alpha$ emission line was also found in the spectrum of GX 339-4
during the bright phase of its 2002-2003 outburst, in a ``very high''
state. As Miller (2007) pointed out, this observation is of interest
for understanding the relativistic iron lines observed in Seyfert 1
AGN. We discuss these observations in connection with the detection
of broad iron K and L line emission in the narrow-line galaxy 1H0707-495.

The existence (at least for some time) of a weak disk in the innermost 
region around stellar mass black holes during the hard spectral state 
strengthens the expectation
to find the same phenomenon for accreting supermassive black
holes as well. Seyfert 1 AGN are the class of supermassive black holes that
offer the best view of the innermost accretion region. Tanaka et
al. (1995) observed the first asymmetric 
disk line profile in the Seyfert 1 AGN MCG-6-30-15 using the ASCA/SIS. Now 
{\it {XMM-Newton, Chandra}} and {\it{ Suzaku}} allow us to detect and
measure relativistic disk lines in Seyferts. The Fe K-shell emission
lines are the strongest X-ray emission lines both in AGN and
X-ray binaries. In a detailed review of relativistic X-ray emission
lines from inner accretion disks around 
black holes Miller (2007) discusses observational and theoretical
developments. The review of Nandra et
al. (2007) focuses on broad iron lines in Seyfert
galaxies observed by {\it{XMM-Newton}} and raises the question how
frequent this broadening is, whether it originates in an accretion
disk, and how robust the evidence for an accretion disk
is. At first glance it is not clear whether the lines originate in a
weak disk below a prominent hard coronal X-ray flux or are connected with 
accretion via an untruncated disk during the ``very high'' state.

\begin{figure}
\begin{center}
\includegraphics[width=8.8cm]{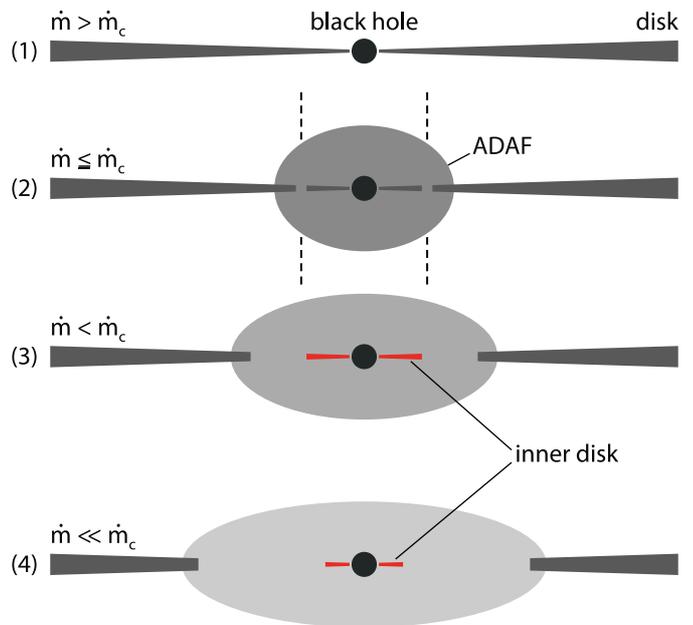}
\caption{\label{schematic} Change of the geometry of the accretion
  flow with decreasing mass accretion rate scaled to the
  Eddington rate $\dot m$, ($\dot m_c$ critical rate for which the
  state transition happens): (1) soft state, (2) transition to the
  hard state begins, formation of a gap where evaporation is most
  effective (3-4) hard states with disk truncation receding outward,
  the weak inner disk disappears (Fig.1, Meyer-Hofmeister et al.2009).}
\end{center}
\end{figure}

The aim of our paper is to analyze under which circumstances the broad
iron emission lines in Seyfert galaxies may originate in a weak disk
in the innermost region around the black hole,
similar to the situation of stellar mass black holes (and neutron
stars). In Section 2 we refer to the process 
of disk evaporation, which for low accretion rates leads to the
truncation of the standard Shakura-Sunyaev disk at a certain radius. In
Section 3 we discuss how changes of the mass flow rate result
in distinct spectral states for both LMXBs and AGN. The mass
flow toward the inner regions can be modulated by the
ionization instability and magnetorotational and gravitational
instabilities (Siemiginowska et al. 1996, Menou \& Quataert
2001). In
the Appendix we discuss how the ionization instability can modulate
the mass flow toward the inner regions. As a consequence of the disk 
evaporation process a  new picture arises for the presence of these
disk instabilities. 

These mass flow variations can cause changes between hard and soft
spectral states and lead to the temporary existence of a weak inner disk.
We discuss in Section 4 for which rates the appearance of an inner disk
can be expected as a transient phenomenon in Seyfert galaxies. In
Section 5 we derive 
accretion rates from observations for the best candidates of broad iron
emission lines in the samples of Seyfert galaxies of Miller (2007) and
Nandra et al. (2007). In Section 6 we discuss a different
accretion flow geometry: broad emission lines from an untruncated disk
in bright sources in the ``very high'' state. 

We note that a different picture is also
discussed for the accretion flow geometry, an always present
accretion disk together with a jet. For supermassive black holes,
Sgr A* and low-luminosity AGN such 
a jet-dominated situation was suggested (Falcke \& Biermann 1999, 
Falcke \& Markoff 2000, Falcke et al. 2000) as an alternative to the ADAF
solution (Narayan et al. 1998, Di Matteo et al. 2000, 2003). 
The iron emission lines
could be caused by illumination from a non-thermal jet.

\section{Evaporation leading to a truncated disk}
The process of evaporation of matter from an accretion disk via a
siphon flow to a hot corona/ADAF was generally proposed for disks surrounding a
compact object (Meyer \& Meyer-Hofmeister 1994). For the interaction between
accretion disk and corona/ADAF around compact objects of different
mass the physics is the same from neutron stars and black holes in
LMXBs to supermassive black holes in AGN (Meyer et al. 2000, Liu et
al. 2002). It is always the interaction
between a coronal hot gas and a cool disk below: frictional heat released in the
corona flows down into cooler and denser transition layers; there it
is radiated away if the density is sufficiently high; if the density is
too low, cool matter is heated and evaporates into the corona until
an equilibrium density is established. Compton cooling adds to thermal
conduction (Liu et al. 2007). Modeling allows us to 
investigate the effects of the chosen parameters, e.g. the viscosity
value, or the inclusion of a magnetic pressure (Qiao \& Liu 2010) and
to take into account hard and soft
irradiation. Exact values for disk truncation and transition
luminosity are difficult to obtain because of this freedom in the
parameter choice and the limitations of a one-zone model used for
describing the complex situation. 

The surface temperatures in the disks
around massive black holes do not change the situation, as was
questioned in the work by McHardy (2009), because disk surface
temperatures are much below the temperatures of the decisive radiating 
layers at the bottom of the coronae. Evaporation and disk
truncation in AGN were investigated by Liu \& 
Meyer-Hofmeister (2001). In recent work Liu \& Taam (2009) suggest
that the existence/non-existence  of the so-called broad line region of
type 1 AGN is related to disk truncation.

\section{Distinct spectral states of LMXBs - analogous situation in AGN?}

For LMXBs a wealth of information on spectral states and transitions
between them is available, because of the short timescales 
(shorter than for AGN - the
dynamical time scales with the central mass, a factor of $10^6$ to
$10^8$ difference) and the large number of objects
in our vicinity. Chen et al. (1997) collected all 
available data at that time, 66 recorded outbursts already. The two
review papers of McClintock \& Remillard (2006) and Remillard \&
McClintock (2006) give a compilation of the large number of observations and
their theoretical interpretations. Accretion flows in black hole X-ray
binaries and neutron star binaries are investigated by Done et
al. (2007). The recent study by Dunn et al. (2010y) includes additional
recent observations.

What is the situation for AGN? Theoretically we expect the same 
properties of the accretion flow, distinct states in
response to changes of the mass accretion rate, for all compact objects of 
different mass (Narayan et al. 1998). Observations for AGN
seem to confirm this situation. In analogy to LMXBs in hard spectral
state, low-luminosity accretion and truncated
disks were found for AGN already many years ago (Narayan 2005). 
Ho (2009a) recently concluded that
the massive black holes in most nearby galaxies reside in a low or
quiescent state. 

For LMXBs we know the clear change of spectra during
the hard/soft and soft/hard transitions of Cyg X-1 in 1996 (Esin et
al. 1998), the change of the photon index $\Gamma$. But for AGN the 
situation is not as clear, we can only draw some parallels from average spectral
energy distributions. Vasudevan \& Fabian (2007) compared the slopes 
in the spectral energy distribution for sources in the ranges 
$0.0012<L_{\rm{bol}}/L_{\rm{Edd}}<0.032$ and 
$0.27<L_{\rm{bol}}/L_{\rm{Edd}}<2.7$ and found a harder spectral shape
for the lower luminosities, and softer X-ray spectra for the higher
luminosities. Done \& Gierli\'nski (2005) argued that complex
absorption from a disk wind might affect the spectra, so that these could
appear as being harder. Shemmer et al. (2006) studied the hard X-ray spectral
slope as an accretion rate indicator for luminous AGN and found a 
dependence on the accretion rate for their sample.

Similarity between accretion in galactic and
supermassive black holes is also found by comparing 
X-ray and optical variability. Measuring the X-ray variability power
spectral density (PSD) of six Seyfert 1 galaxies Markowitz et al. (2003)
pointed out a physical similarity with X-ray binaries. But Uttley \& McHardy
(2005) found, e.g. for NGC 3227, an intrinsically hard X-ray
spectrum (photon index $\Gamma\sim 1.6)$, yet a broad-band X-ray
variability PSD reminiscent of black hole X-ray binaries in the
high/soft state, and argued that the current nomenclature for the
various states may be inappropriate. More Seyfert galaxies were
investigated by Uttley \& McHardy (2005), McHardy et
al. (2005) and Ar\'evalo et al. (2006).

Because mass flow changes are a basic element in the consideration of
spectral states in AGN as well, we add to our analysis a discussion of
disk instabilities in the Appendix, which can produce these changes and
trigger spectral transitions. We show that a new picture of the
possible importance of the ionization instability arises from the
truncation of inner disk regions, which can lead to an elimination of
the instability. But the instability is expected to be present for
black hole masses of $10^7M_\odot$ to about $10^8M_\odot$. 
 
\section { A cool disk in the innermost region ?}

 \subsection {The re-condensation model}

Observations of LMXBs indicate cool gas in
the neighborhood of the accreting black hole. Most observations were
made at a time shortly
after the soft/hard spectral state transition, which points to an 
intermediate state connected with the transition process.

 During phases of decreasing mass flow rate, as in outburst decline, 
a gap opens between the outer disk and the inner region at the 
distance where the evaporation is most efficient, which is filled by
an ADAF (Fig.1). Within the short diffusion time
the inner disk would disappear, if it were not sustained by a continuous
supply of gas from the ADAF above. The observations indicate that the
inner disk can survive. The re-condensation model (Liu et al. 2006, 
Meyer et al. 2007) allows us to understand the interaction:
the ADAF is affected by thermal conduction to a cool accretion disk underneath.
Such a model takes into
account the exchange of energy and mass between the tenuous,
inefficiently radiating, hot two-temperature corona and the uppermost
layers of the disk. The re-condensation rate becomes a function of the mass 
flow rate in the ADAF, $\dot m_{\rm{ADAF}}$ (in Eddington accretion
rate units), and 
the ratio of the distance to the ISCO and the distance of the outer edge 
of the inner disk,  $R_c$, which depends on $\dot m_{\rm{ADAF}}$.

\begin{equation}
\dot m_{\rm{cond}} / \dot m_{\rm{ADAF}} = 3.23\times 10^{-3} \alpha
^{-7} \dot m_{\rm{ADAF}}^2 f(R_{\rm{ISCO}}/R_c)
\end{equation} 
 with the function 
\begin{equation}
f(x)=1-6x^{1/2} +5x^{3/5}.
\end{equation} 

The process works best at distances
from 30 to 100 Schwarzschild radii, and suitable condensation rates are typically
10\% of $\dot m_{\rm{ADAF}}$
(strongly dependent on the  viscosity parameter $\alpha$). All condensated
gas flows inward to the ISCO. As $\dot m_{\rm{ADAF}}$ decreases, the inner
disk becomes smaller and the re-condensation rate decreases. This
inner disk can exist for a long time during the low/hard state, 
as long as the accretion rate is not too low, $\leq \dot m \approx 10^{-3}$ 
 (Taam et al. 2008), and could still be present during
the rise to the next outburst.

The model was applied to X-ray observations of the black hole
sources GX 339-4 and SWIFT J1753 and is able to explain the effective 
temperature of the thermal component
(Liu et al. 2007). Compton cooling  was found to be an important
element in the re-condensation process (Taam et al. 2008).

 The physics of the re-condensation process is the same for stellar mass
and supermassive black holes. The disks in AGN have lower
temperatures, thermal conduction occurs as in LMXB disks. The
diffusion time for AGN disks close to the black hole is also short
compared to the time within which the mass flow rate from the outer
regions can change. This means that re-condensation is needed in AGN 
as well to sustain a cool innermost disk underneath the ADAF.

\subsection {Cool disks around stellar mass black holes. The observational
  evidence}

For investigating inner disks in AGN, the observations of 
stellar-mass black hole sources are of interest.
Besides the mostly cited observations of GX 339-4 and SWIFT J1753.5-0127
(Miller et al. 2006a, 2006b, Reis et al. 2008,
Tomsick et al. 2008), indications for the presence of weak inner disks 
are also found 
in several other sources (for the strength of the thermal components
see Meyer-Hofmeister 2009), very recently also in the new X-ray
transient XTE J1652-453 (Hiemstra et al. 2010). 
These thermal components were found in connection with outbursts, in
outburst decline, usually during the brightest phases of the hard
state (with the exception of Cyg X-1). Recently Reis et al. (2010)
presented a study of eight stellar-mass black hole sources in the 
low/hard state and confirmed the existence
of a thermal component for all of them, with iron K$\alpha$ line
emission in half of the sources.
All observations of thermal components close to the ISCO 
document the presence of cool material close to the black hole
(i.e. for dynamical reasons an inner disk) during the hard state. 
Unfortunately we cannot yet derive from this whether such an inner
disk continues all the way out or if a gap exists at the distance
where (according to the modeling) evaporation is most efficient.

Interestingly, a skewed relativistic Fe K$\alpha$ emission line
  was also observed for XTE J1650-500 and for GX 339-4
in ``very high `` state (Miller et al. 2002, 2004), probably originating in
a different accretion flow geometry for a high accretion rate, an
untruncated disk in a very bright state.

\subsection {Mass flow rates appropriate for inner disks around
  supermassive black holes}
For AGN the situation should be similar to that for stellar mass
black holes (for a definition of states see Remillard 
\& McClintock 2006): (1) a low/hard state, an ADAF in the inner region,
accretion rates below a few percent to 0.1 $\dot
M_{\rm{Eddington}}$ (spectrum hard); (2) a standard high/soft state
(=''thermal''), disk inward to the ISCO, accretion rates about 0.05 to 0.5 $\dot
M_{\rm{Eddington}}$ (spectrum soft); (3) a very high state (``steep
power law''), disk inward to the ISCO and an additional
coronal layer above the disk, accretion rates just sub-Eddington 
(spectrum soft + hard component). During a decrease from the high/soft
state to the low/hard state for some time a weak inner disk
can be present (spectrum hard + weak thermal component).

We here focus on the situation shortly after the transition from
the soft to the hard state, with an additional
weak inner disk, as a relict of a former disk that reached inward to the
ISCO in a preceding high/soft state. This means the accretion rates should lie
between $\dot m_{\rm{crit}}\approx 0.05 $ to $0.1$ and about
$\frac{1}{10}$ of this rate (based on the evidence of such inner
 disks in LMXBs). This predicts accretion rates appropriate to allow inner
 disks in Seyfert 1 galaxies of $0.1M_\odot /
{\rm{yr}}$ for a $10^8 M_\odot$ black hole, one tenth of this rate for
a $10^7M_\odot $ black holes (in our discussion of mass flow rate
changes caused by the ionization instability these rates correspond to
the dashed lines in Fig.B1, Appendix).

\section{Broad iron lines in Seyfert 1 galaxies as
  indicators of an inner accretion disk}
\subsection{Observations}

Our search for inner disks in AGN was motivated by the observation of
broad iron lines in Seyfert galaxies, as described in the review of
relativistic disk lines by Miller (2007) and confirmed by
recent observations with {\it{Chandra, XMM-Newton}} and {\it{Suzaku}},
and also discussed in the analysis of spectra by Nandra et
al. (2007). These emission lines are the most
obvious reaction of an accretion disk to irradiation by an external
source of hard X-rays, which is true whenever a corona lies above a
disk. The first clear example of relativistic dynamics affecting the line
shape, that is a line originating in the innermost region, was found
in the Seyfert 1 AGN MCG-6-30-15 by Tanaka et al. (1995), at that time
using ASCA/SIS.  

Miller (2007) compares how frequent detections of relativistic
disk lines in LMXBs and Seyfert 1 AGN are: (1) in 75-85\% of the stellar
sources relativistic disk lines were found. (2) For the
sample of AGN with broad iron emission lines chosen by Nandra et
al. (2007) the authors found that 45\% of the sources are best fit
with a relativistically blurred reflection model. But the comparison is
difficult. For LMXBs inner weak disks were preferentially observed
after the soft/hard
transition, during a phase of accretion rate decrease, an only transient
phenomenon. That phase lasts only a short time of the full outburst
cycles, including quiescence.  The detections in AGN refers to the
sample of sources selected because of their iron line emission. 

\subsection{The accretion rates of the best candidates}

\begin{figure}
\begin{center}
\includegraphics[width=8.8cm]{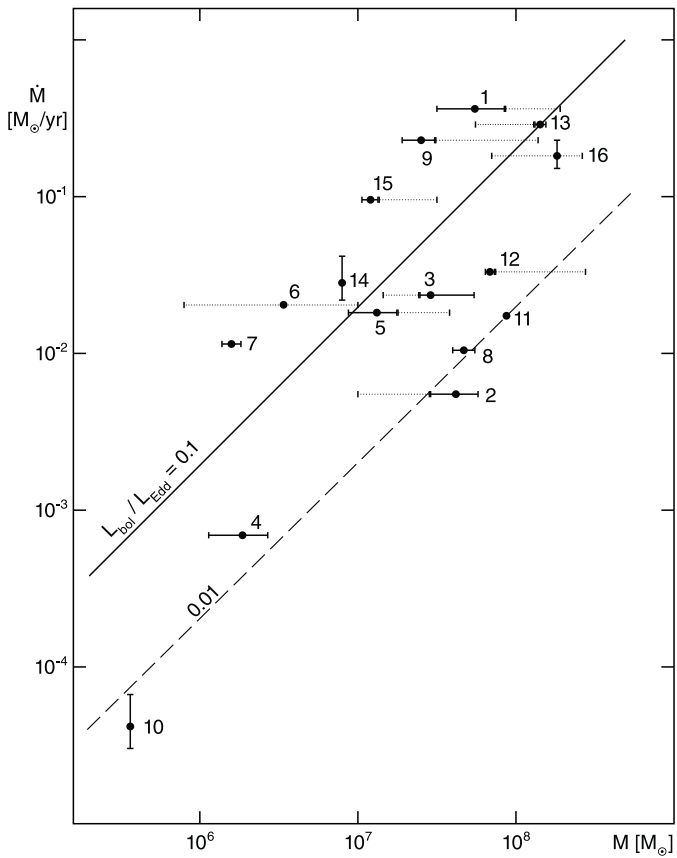}
\caption{\label{}Accretion rates of Seyfert galaxies with strong 
  iron emission lines from inner accretion disks revealed in deep 
  {\it{Chandra,
  XMM-Newton}}, and {\it{Suzaku}} observations (Miller 2007,
  Nandra et al. 2007); numbers in Table 1. Solid line:
  critical accretion rate corresponding to the soft/hard state
  transition; long-dashed line: 1/10  of critical rate; theory predicts
  the possible existence of an inner disk for accretion rates
  between these two lines. Horizontal error bars, solid lines: from Peterson et
al. (2004), Nandra (2006); dotted lines: from Niko{\l}ajuk et
al. (2006), Iwasawa et al. (2004), Markowitz et al. (2007), Boller et
al. (2007); vertical error bars for 2-10 keV
luminosities and bolometric corrections given for sources 10, 14 and 16
(Vasudevan and Fabian 2007); not indicated for sources 1-5, 8, 9, 12,
13, and 15 (Vasudevan and Fabian 2009) because the errors in the 2-10 keV
luminosities are small (see their Table 2), errors in bolometric
corrections only a few percent. But see Sect. 5.2 for the general 
uncertainty of unscaled X-ray luminosities.$\rangle$}
\end{center}

\end{figure}

 We take the information from two samples of Seyfert galaxies. Group
(1) are sources with ``very strong relativistic disk line detections'',
Tier 1 of the census of Miller (2007). Group (2) are sources with
``acceptable model fits including a broad Gaussian emission
line component or a relativistically blurred line added to the
spectrum'' from the analysis of Nandra et al. (2007). We have not
tried to exclude sources that might be in a high/soft spectral state.

Considering our sample of sources some special features need be commented.
For some sources in group (2) differing results were found in multiple 
observations. The authors argued that non-detection of 
broad emission lines may be due to a low signal-to-noise ratio. 
The Nandra sample contains most of the sources in Millers Tier 1, his
best candidates, but the judgment on the 
emission lines is not always the same: for NGC 3783 and MCG-6-30-15
the model fits of Nandra et al. were classified as poor. We kept these
sources in our chosen sample. Despite its poor fit we included Ark 120 
(sample of Nandra) in our sample because it is in Tier 3 of
Miller (sources with detections that need to be confirmed and investigated
more deeply). Because of heavy contamination with light from the
galaxy (according to Vasudevan \& Fabian 2007), IC 4329A was not
included. 

For some sources a different X-ray luminosity is found from
observations at different time, e.g. for NGC 4151 a change in $L_X$
from $10^{42}$ to about $10^{42.8}$ erg/s between December 2000 and May
2003. For this observation in May 2003 from spectral fits Nandra et
al. (2007) derived the value $10^{42.8}$erg/s, Vasudevan \& Fabian
(2009) the value $10^{42.64}$ erg/s. For NGC 3783 the luminosity found from
observations on 2001 December 19 is 50\% higher than that of December
17 (the difference maybe caused by an outflowing warm absorber). As the
derivation of luminosity values for the same observations by different
authors shows, the results of spectral fits occasionally 
differ by about 30\% (in NGC 3783 and NGC 3516) and even more for NGC 4051 as 
observed in May 2001.

The samples of Miller (2007) and Nandra et al. (2007) were also used in a
recent analysis of Brenneman \& Reynolds (2009) who selected eight
candidates to investigate the relativistic broadening of iron emission
lines for spin diagnostics. In half of the sources they found 
an improvement in their goodness-of-fit parameters when
relativistic effects from an inner disk were taken into account.

The sources we finally consider are 3C 120, NGC 3516, NGC 3783,
NGC 4051, NGC 4151, Mrk 766, and MCG-6-30-15 from Miller (2007) and in
addition Mrk 590, Mrk 110, NGC 4395, NGC 5506, NGC 5548, Mrk 509, Ark
564, NGC 7469, and Ark 120 from Nandra et al. (2007). To clarify whether
the broad iron emission lines of the sources in this sample could
originate in a weak inner accretion disk that is 
surrounded by the ADAF carrying the main part of the accretion flow, we
determine $L_{\rm{bol}}/L_{\rm{Edd}}$ and take this ratio as an
approximate measure for the scaled accretion rate $ \dot M/\dot
M_{\rm{Edd}}$. 

For this evaluation the bolometric luminosities (from
X-ray luminosities and X-ray bolometric corrections) and
reliable black hole masses are needed. Recently Vasudevan \& Fabian 
(2007, 2009) determined X-ray bolometric corrections using spectral energy 
distributions (SED) in different wavebands (possible errors are
between 10 and 100 \% in the first work, but only 1 to 10\% in the
second). We take
the values for X-ray luminosity, black hole mass, bolometric
correction and $L_{\rm{bol}}/L_{\rm{Edd}}$ from these investigations
(where available the newer values). For sources not included in this 
investigation we take the X-ray luminosity from Nandra et
al. (2007), together with a constant bolometric correction as used by 
Ho (2009a), and the black hole mass from O'Neill et al. (2005). For a few
sources in our sample X-ray luminosities are also listed in the
investigation of Ho et al. (2009b), estimates differ by up to a factor
of three (even larger for NGC 4051). In Table 1 we list the derived
values. For 10 sources the Eddington-scaled accretion rates are 
$\leq$0.1. This value can be considered an upper limit for the
transition between hard and soft spectral state). (Because the 
theoretically derived transition luminosity depends on parameters whose exact values are not determined
by the simple modeling, but have to be assumed, one has to resort to
observations for estimating the critical values; see also the values
for the transition luminosity derived by Gierli\'nski \& Newton (2006)). For six
sources rates clearly above 0.1 are found (possibly erroneously high values, see error bars).

The uncertainty of black hole masses affects the values of the
Eddington-scaled luminosity. The values of black hole masses in the
literature (also in Vasudevan \& Fabian 2007, 2009) largely go back to the
compilation of Peterson et
al. (2004), based on broad emission-line reverberation-mapping data
with a typical uncertainty of 33\%. Comparing with black hole masses 
evaluated by other methods, e.g. the often used relation between black
hole mass and central stellar velocity dispersion, a similar uncertainty
is found. Niko{\l}ajuk et al. (2006) compared black hole masses
determined from reverberation and the X-ray excess variance method and
found large differences. For sources included in the recent 
catalog of central stellar velocity dispersions of nearby galaxies  
(Ho et al. 2009b), the black hole masses
evaluated from the velocity dispersion values using the relation determined by
Tremaine (2002) differ by a factor of $\le$3.

\subsection {Interpretation of the accretion rates}

For Seyfert galaxies NGC 3516, NGC 3783, NGC 4051, NGC 4151, Mrk 590, 
Ark 120, NGC 4395, NGC 4593, NGC 5506, and NGC 5548 the 
Eddington-scaled accretion rates are below 0.1. This can be theoretically 
interpreted as a low spectral state, with an ADAF in the innermost
region. The disk from which the emission lines come should then be a
weak inner disk, separated by a gap from the outer truncated accretion disk, 
an accretion flow geometry similar to that in LMXBs. 

In Fig.2 we show the distribution of obtained accretion rates, given in
$M_\odot/{\rm{yr}}$. The bolometric luminosities vary from 
$4\, 10^{-5}$ to 0.367 $M_\odot/{\rm{yr}}$. Error bars owing to
uncertainty in black hole mass are widely given in the literature and
are indicated
in our figure. Errors in unscaled X-ray luminosity are difficult to
judge (compare our discussion in Sect. 5.2). According to Vasudevan
and Fabian (2009, Table 2) these errors are small. Error
bars owing to uncertainty in bolometric correction 
are small for the sources examined in the work of Vasudevan
and Fabian (2009), a few percent only. 

For six sources rates above 0.1 are found. This might be owing to errors
because uncertainties enter in several ways. In the picture 
of a thin disk reaching all the way inward or an inner ADAF plus a truncated
outer disk, depending on the accretion rate, one expects that these sources are in a standard high/soft
spectral state with an untruncated disk reaching inward to the ISCO.
But the photon index $\Gamma$ listed by
Vasudevan\& Fabian (2009, Table 2) for these sources is $\leq$1.95,
or even lower, $\leq$1.5. (For X-ray binaries Remillard \& 
McClintock (2006) chose 1.4 $\leq \Gamma \leq $2.1 for their
definition of the low/hard state.) One of these sources, MCG-6-30-15,
was classified as in high/soft state based on the comparison of the
PSD (Uttley \& McHardy 2005), but the same was found for NGC 4051, for
which a low accretion rate (scaled to the Eddington rate) was found. 
These examples show that it is more difficult to confirm the spectral
state for AGN than for LMXBs.

For LMXBs an interesting observation is that e.g. GX 339-4 at one time was found
in low/hard spectral state and at another time in the very high
state with an extremely skewed
relativistic Fe K$\alpha$ emission line (Miller et al. 2004). 

It is conspicuous that in the distribution of accretion rates of the 
systems with Fe K$\alpha$ emission lines a
dichotomy between low and very high values is
found for LMXBs, but not for Seyfert galaxies
with the same reflection features. For Seyfert galaxies we find
sources with these rates in between the low and very high rates of
LMXBs. Theory would predict 
continuous accretion disks inward toward the ISCO for these sources. If
irradiation seems to be present this would mean that the innermost
coronae in these AGN would be significantly stronger than those in
LMXBs at comparable accretion rates. What could cause these differences 
between AGN and LMXBs? LMXBs are accreting binary systems, AGN accrete
gas that has fallen in at large distances. That could in principle
make a difference in the
accretion of magnetic flux through their accretion disks, and this flux
is possibly closely related to coronal dissipation and jet
formation. These are interesting and challenging questions, but more
theoretical  work and probably also more observations are needed to
make progress.

\begin{table}
\begin{center}
\caption{
Accretion rates deduced for Seyfert 1 galaxies with broad iron
emission lines: [1] $L_X$ 2-10 keV luminosity in units of erg/s,
[2] $M_{\rm{BH}}$ black hole mass in units of $M_\odot$, [3]
bolometric correction $\kappa_{\rm{2-10keV}}=L_{\rm{bol}}/L_X$, 
[4] $L_{\rm{bol}}/L_{\rm{Edd}}$, 
[5] $\dot M$ in $M_\odot/{\rm{yr}}$, $\dot M/ \dot M_{\rm{Edd}} 
\approx L_{\rm{bol}}/L_{\rm{Edd}}$}  
  
\begin{tabular}{llllll}
\hline
&&&&&\\
\  source & log$L_X$ & log$M_{\rm{BH}}$ & $\kappa_{{2-10}\rm{keV}}$
& $\frac{L_{\rm{bol}}}{L_{\rm{Edd}}}$&$\dot M$\\
&&&&&$[M_{\odot}/y]$\\
&[1]&[2]&[3]&[4]&[5]
\\
\hline
&&&&& \\
(1)\ \ 3C 120& 44.0& 7.74&20.6 &0.305\ $\times$&0.367 \\
(2)\ \ NGC 3516& 42.3& 7.63&15.3 &0.006&0.006\\
(3)\ \ NGC 3783& 42.9& 7.47&17.3 & 0.036&0.023\\
(4)\ \ NGC 4051& 40.8& 6.28&67. &0.016&0.001\\
(5)\ \ NGC 4151& 42.8& 7.12&17.4 &0.062&0.002\\ 
(6)\ \ Mrk 766& 42.88& 6.54&15.8 &0.275\ $\times$&0.021\\
(7)\ \ MCG-6-30-15& 42.65& 6.19&15.8 &0.362\ $\times$ &0.012\\
(8)\ \ Mrk 590& 43.0& 7.68&7.0 &0.010&0.011\\
(9)\ \ Mrk 110& 43.9& 7.40&18.4 &0.433\ $\times$& 0.238\\
(10)\ \ NGC 4395& 40.02& 5.56&22.5 &0.005&$4\,10^{-5}$\\
(11)\ \ NGC 5506& 42.80& 7.94&15.8 &0.009&0.017\\
(12)\ \ NGC 5548& 43.3& 7.83&10.1 &0.024&0.035\\
(13)\ \ Mrk 509 & 44.0&8.16& 16.2&0.095&0.301\\
(14)\ \ Ark 564& 43.25& 6.90&9.2 & 0.162\ $\times$&0.028\\
(15)\ \ NGC 7469& 43.2& 7.09&42. &0.369\ $\times$ &0.099\\
(16)\ \ Ark 120& 43.82& 8.27&15.7 &0.044& 0.180\\
\hline 
\end{tabular}
\end{center}

References: for sources 1-5, 8, 9, 12, 13, and 15 Cols. [1]
to [4] from Vasudevan \& Fabian (2009); for sources 10, 14, and 16 
Cols. [1] to [3] from Vasudevan \& Fabian (2007), for
sources 6, 7, and 11 $L_X$ from Nandra et al. (2007), 
$M_{\rm{BH}}$ from O'Neill et al. (2005), assumed constant bolometric correction 
15.8 (Ho 2009a); crosses mark sources with
an Eddington-scaled mass flow rate $\ge 0.1$.

\end{table}

\section{Broad iron emission lines from untruncated disks:
 the narrow-line Seyfert 1 galaxy 1H0707-495}
For 1H0707-495, a narrow-line Seyfert 1 galaxy (NLS1), broad iron K and L
emission lines were recently found (Fabian et al. 2009).
The bright Fe-L emission allows us to detect a reverberation lag of 30s 
between the direct X-ray continuum and its reflection from matter falling
into the black hole, indicating that we obtain information from matter within a
gravitational radius. 

The accretion rate is estimated to be (just) 
sub-Eddington. The observations point to an untruncated disk reaching 
inward to the ISCO and Comptonization of the irradiating X-rays by the
disk (for a discussion of the very
high state see Done, Gierli\'nski, and Kubota (2007)). These observations
prove that Fe K$\alpha$ emission
lines can be present in quite different geometries, always from
disks in the innermost region, but possibly of two different kinds, weak inner
disks in the hard spectral state or strong untruncated disks in a very
high state. Spectral energy distributions from contemporaneous optical, UV and
X-ray observations are needed to distinguish between these spectral
states. 

Some of the Seyfert 1 galaxies investigated for their broad iron
emission lines by Miller (2007) and Nandra et al. (2007) might have
similarity with 1H0707-495, but are not as bright as this
source. Actually Ark 564 is known as a narrow line Seyfert 1
galaxy (Ar\'evalo et al. 2006).

\section{Conclusions}
For LMXBs broad iron emission lines found in an intermediate
hard spectral state were interpreted as originating from a weak
inner disk, which is left over from a disk reaching inward to the
ISCO at earlier time. The accretion flow
geometry in the thin disk+ADAF model scales with the black hole
mass. We discussed whether such a model for the interpretation of iron
emission lines is also applicable to Seyfert galaxies.

We used recent observations and interpretations of
broad, partly relativistically blurred iron emission lines in Seyfert
1 AGN. We considered sources in a hard spectral state. For the
best candidates
in the samples of Miller (2007) and Nandra et al. (2007) we evaluated 
accretion rates (scaled to the Eddington accretion rate) to see
whether these rates point to a hard spectral state. As a limiting
Eddington-scaled upper accretion rate for sources in the hard spectral
state we chose the value 0.1, taking into account the observational
results for the state transition luminosities of LMXBs. It is
difficult to judge the uncertainty of the accretion rates evaluated for the
Seyfert galaxies. Therefore we cannot definitely determine the
spectral state. As dicussed in Sect.3, it is not possible to derive
firm information on the spectral state from the photon index of the 
spectral energy distribution. 

For the majority of the  sources considered here we found
accretion rates $\le$ 0.1 (see Table 1), agreeing with 
the theoretical expectation of a weak inner accretion
disk below the ADAF during the low/hard state for AGN as well. 
The appearance of iron K$\alpha$ lines signifies the presence of
cool material close to the accretng black hole, irradiated by a
power-law continuum from a corona above. In this state the inner disk
carries only a fraction of the total accretion flow, the major part
flows through the corona and provides the hard luminosity. This
situation is desribed in the
modeling of the ``truncated disk + corona + re-condensed inner disk'' 
geometry. Disk truncation is an inherent feature of disk
evaporation that explains the appearance of the hot, optically thin
flow for low accretion rates (and also the hysteresis in transition
luminosity between soft and hard spectral states).

Besides the Seyferts with accretion rates below the critical value
there is a significant number of sources in our sample (40\%) that
show accretion rates above this limit, though not as high as in some
LMXBs in high state with iron emission lines (a significant fraction of
the Eddington value). These LMXB sources probably are related to the
so-called ``very high'' state, where the disk material is 
illuminated by a strong corona on top of an inner disk
(Done, Gierli\'nski and Kubota 2007). In the LMXB case a
clear gap exists in the accretion rate distribution between the two
groups with low and high rates, no sources with rates slightly above
0.1 are known. The galactic sources
XTE J1650-500 and GX 339-4 were observed in such a very high state, and
extremely skewed relativistic Fe K$\alpha$ emission lines were
detected (Miller et al. 2002a, Miller et al. 2004). Miller
(2007) pointed out that the spectral and variability phenomena then closely
resemble the behavior seen in Seyfert AGN. The
narrow-line Seyfert 1 galaxy 1H0707-495 seems to be a
particularly bright source of this kind, with an accretion rate just
below the Eddington rate.

The Seyfert galaxies with accretion rates somewhat about 
the critical value (with no analogous stellar sources) deserve particular
attension because the existence of 
these sources might indicate a difference between the accretion flow 
geometry around stellar mass and supermassive black holes. Theory
predicts an untruncated disk for sources with accretion
rates above 0.1. The reflection features then indicate the presence
of significantly stronger coronae at comparable accretion rates in
these AGN than in LMXBs. If strong magnetic flux is the cause of this
phenomenon, these coronae might well also be the base of a jet issuing
from them. Such differences between accretion in stellar mass and 
supermassive black holes certainly deserve further work.

Weak inner disks below an ADAF-type accretion flow in the inner region
of an accreting black hole can be considered as the natural remainder
of an originally
standard Shakura-Sunyaev disk that had reached inward all the way to the ISCO
at high accretion rates and has now, when the accretion rate has
dropped below its critical value, become truncated. This
could therefore be a hint to a higher accretion
rate in the past. Moderate accretion rate variations caused by the
ionization instability in AGN disks could well allow the inner disk to 
exist for long times (as in the case of the galactic black hole binary
Cyg X-1). 
Seyfert 1 galaxies with $10^7$ to $10^8 M_\odot$ black holes and 
accretion rates of about 0.01 to 0.1$M_\odot/{\rm{yr}}$ seem specially 
suited to display this weak inner disk accretion flow geometry.

\begin{acknowledgements}
For helpful discussions and suggestions we thank Bifang Liu.

\end{acknowledgements}

{}

\appendix

\section{The ionization instability in disks around stellar mass 
and supermassive black holes} 
Accretion disks are thermally and viscously unstable at radii where
the central temperature lies between about a few times $10^3$ K and
$10^4$ K. As a result the disk structure alternates between 
a hot, ionized structure and a cool non-ionized structure, corresponding to
a high (outburst) or low  (quiescence) mass flow rate in the disk. This 
instability is the cause of the dwarf nova outbursts (Meyer \& 
Meyer-Hofmeister 1981, Smak 1984, for a review of model calculations
see Cannizzo 1993). The ionization instability is also the cause of
the outbursts of soft X-ray transients (Dubus et al. 2001 and
references therein). 

The application of the disk instability model to AGN (Lin \& Shields
1986, Cannizzo 1992, Siemiginowska et al. 1996) is important in
the framework of our analysis.

In their detailed modeling of the thermal-viscous 
ionization instability in AGN disks, Siemiginowska et al. (1996) had
calculated the cyclic luminosity variation. Menou \& Quataert (2001)
found that in contrast to the situation in dwarf novae and soft X-ray 
transients, the gas remains well coupled to the magnetic field even on
the cold branch of the limit cycle and consequently the parameter describing the
viscosity should remain the same in
hot and cold state. Siemiginowska et al. (1996) had evaluated the
lightcurves for this case as well. Then the luminosity changes  
during the outburst cycle are not large, rather a ``flickering''
appears. In Fig. B1 we show the effective temperature in disks around a
supermassive black hole of $10^7$, $10^8$ or $10^9$ solar masses 
for a variety of mass accretion rates $\dot M$ as a function of the 
distance $R$ 

\begin{equation}
T_{\rm{eff}}=\frac{3}{8 \pi \sigma}\frac{GM \dot M}
{ R^3}(1-\sqrt{R_{\rm{ISCO}}/R}),
\end{equation} 

with $G$ the gravitational constant, $M$ the mass of black hole,
  $\sigma$  the Stefan-Boltzman constant, 
  $R_{\rm{ISCO}}$ the distance of the innermost stable 
  orbit = 3 $R_S$ (non-rotating black hole), $R_S=2GM/c^2$
  the Schwarzschild radius, and $c$ the velocity of light. The occurrence of the
ionization instability depends on the midplane temperatures,
which are typically double the effective temperature. To show the zones affected
by the ionization instability in AGN we take the results from the
investigation of Siemiginowska et al. (1996,  Fig.1):  
 $3.1\leq   log T_{\rm{eff}} \leq 3.6 $ (same viscosity parameter in hot and cold
state, otherwise the instability strip reaches down to lower temperatures).

\begin{figure*}
\begin {center}
\includegraphics[width=16.cm]{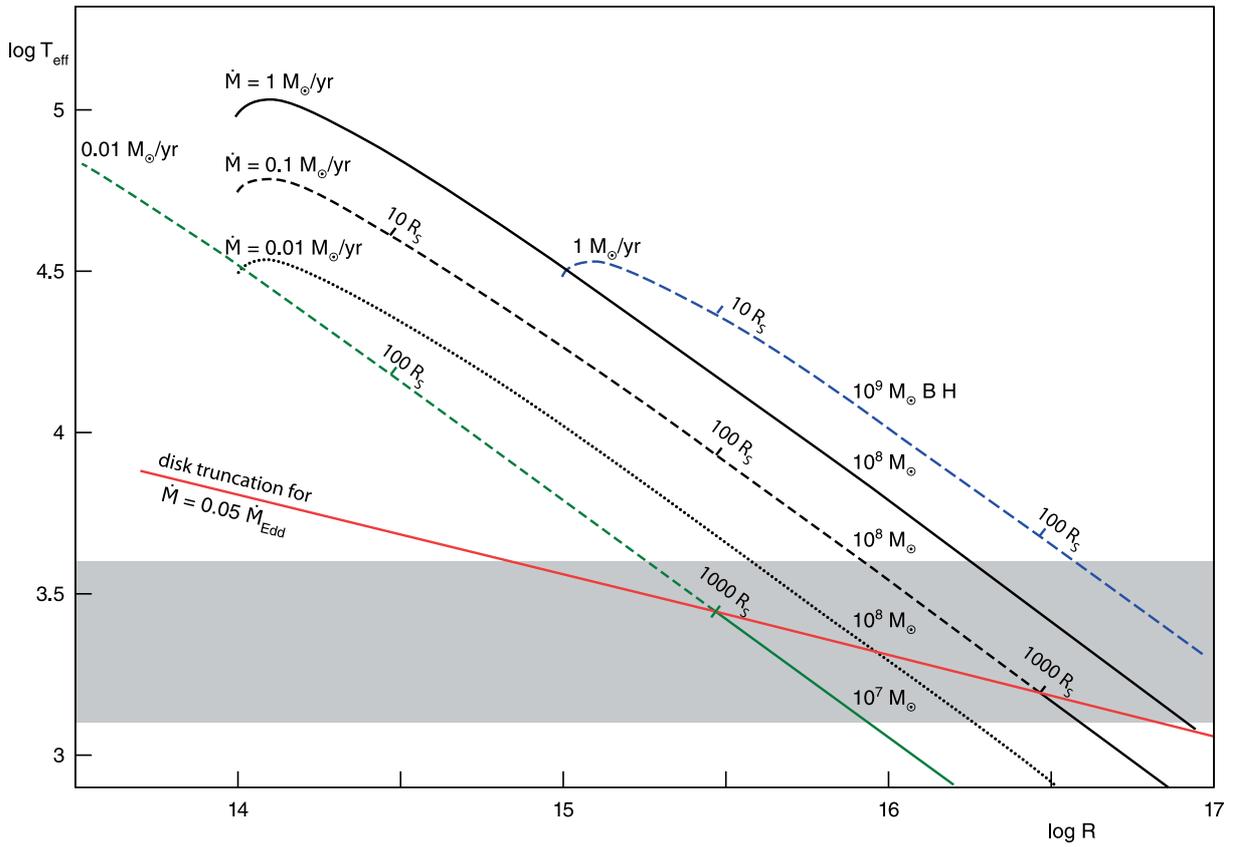}
\caption{\label{schematic} 
Effective temperatures of the disk,
  ionization instability strip (grey area), and disk truncation (inside
  ADAF). Truncation depends on the accretion rate (compare Fig. 1), in
  the figure solid lines for thin disk, dashed and dotted lines for ADAF;
  situation for a $10^8M_\odot$ BH: with $\dot M=1M_\odot/{\rm{yr}}$ disk
  reaches inward to the last
  stable orbit (solid line); for $0.01M_\odot/{\rm{yr}}$ the disk is truncated 
  far out, beyond the range shown, inside ADAF only
  (dotted line); in between these two rates lies the critical 
  accretion rate for spectral transition, of interest here,
  $0.1M_\odot/{\rm{yr}}= 0.05 \dot  M_{\rm{Edd}}$ with disk truncation at
  about 1000$R_{\rm {S}}$ ($R_{\rm{S}}$ Schwarzschild radius) 
  (dashed line/solid line); 
  For $10^9M_\odot$ and $10^7M_\odot$ BHs: only curves
  for the critical accretion rate are shown.   
  } 
\end {center}
\end{figure*}

Menou \& Quataert (2001) analyzed the interplay between
the various instabilities in disks around supermassive black holes. They 
pointed out that in low-luminosity AGN with $\dot M \le 10^{-2}M_\odot /{\rm{yr}}$
the disk is globally gravitationally stable, but in luminous AGN with $\dot M
\ge 10^{-2}M_\odot / {\rm{yr}}$ the disk can be globally gravitationally unstable
except at small radii. We are interested in disks with moderate 
accretion rates and limit our investigation to mass flow
variations caused by the ionization instability.

\section{A new aspect of AGN disk instabilities as a consequence of 
disk truncation}

Disk truncation as an additional feature in the accretion flow
geometry can eliminate the occurrence of the
ionization instability. Evaporation (discussed in Sect.2) is expected
to lead to the disappearance of inner regions, if the accretion rate in
Eddington units lies below the critical value $\dot
m_{\rm{crit}}$. The truncation then occurs at a distance of $R \ge
300 R_S$. For lower accretion rates the truncation might be far out,
at thousands of Schwarzschild radii, e.g. in
low-luminosity AGN (see Narayan 2005, Fig. 3).

 In Fig. B1 we show the situation for different black hole masses and
 different accretion rates. For a 
$10^8 M_\odot$ black hole: (1) for a high accretion rate $\dot
 M=1M_\odot/{\rm{yr}}$ the disk reaches inward to the ISCO; (2) for a
 low accretion rate $\dot M=10^{-2} M_\odot/{\rm{yr}}$ the disk is
 truncated at about $10^{17.1}$ cm (Meyer-Hofmeister \& Meyer 2006), 
far outside the instability strip, (3) for the rate  $\dot
M=10^{-1}M_\odot/{\rm{yr}} \approx \dot m_{\rm{crit}}\approx 0.045$,
which is the critical rate for the spectral state transition 
where we expect a  disk truncation at a distance of about 1000 $R_S$, lying
within the instability strip. For $10^7 M_\odot$ and $10^9 M_\odot$ 
black holes the temperature distribution is only shown for the
critical accretion rate.

As a consequence of disk truncation the unstable regions are partly
eliminated, as can be seen from Fig. B1. For instance an unstable
region as described by Menou \& Quataert (2001, Fig. 1 ) could not be 
present at the distance of 30$R_S$ from a $10^7 M_\odot$ black hole.

For the question whether a weak inner disk could be expected
(accretion rate close to the critical rate for state transition, 
dashed lines in Fig. B1), two
conclusions can be drawn: (1) the ``flickering'' (instead of large
luminosity variation) supports the existence of an inner disk if the
accretion rate is close to the critical one for state transition; (2)
for these rates (dashed lines in Fig. B1) for black hole masses of 
$10^7 M_\odot$ to $10^8 M_\odot$ the disk instability can be present,
for higher masses it may be eliminated, that is, the mass flow can be
stable.

For large black holes of $10^9M_\odot$ the ionization instability
may occur as a special feature in a weak inner accretion disks, if
the temperature there does not allow a stable structure.

\end{document}